\documentclass[useAMS,usenatbib]{mn2e}
\usepackage{natbib}
\usepackage{amsmath,amstext}
\usepackage{graphicx} 
\newcommand{\ci}{C\,{\sc i}}
\newcommand{\oi}{O\,{\sc i}}
\newcommand{\tii}{Ti\,{\sc i}}
\newcommand{\tiii}{Ti\,{\sc ii}}

\newcommand{\Stromgren}{Str\"omgren\ }

\usepackage{amssymb}

\begin{document}

\title[MPs in NGC 6553]{Two Groups of Red Giants with Distinct Chemical Abundances in the Bulge Globular Cluster NGC 6553 Through the Eyes of APOGEE}
\author[Tang et al.]{\parbox{\textwidth}{Baitian Tang$^{1}$\thanks{E-mail:btang@astro-udec.cl},
Roger E. Cohen$^{1}$,
Doug Geisler$^{1}$,
Ricardo Schiavon$^{2}$,
Steven R. Majewski$^{3}$,
Sandro Villanova$^{1}$,
Ricardo Carrera$^{4}$,
Olga Zamora$^{4}$,
D. A. Garcia-Hernandez$^{4}$,
Matthew Shetrone$^{5}$,
Peter Frinchaboy$^{6}$,
Andres Meza$^{7}$,
J. G. Fern\'andez-Trincado$^{8}$,
Ricardo R. Mu\~noz$^{9}$,
Chien-Cheng Lin$^{10}$,
Richard R. Lane$^{11}$,
Christian Nitschelm$^{12}$,
Kaike Pan$^{13}$,
Dmitry Bizyaev$^{13,14}$,
Daniel Oravetz$^{13}$,
Audrey Simmons$^{13}$}\\
 \\$^1$Departamento de Astronom\'{i}a, Casilla 160-C, Universidad de Concepci\'{o}n, Concepci\'{o}n, Chile
 \\$^2$Astrophysics Research Institute, Liverpool John Moores University, Liverpool, United Kingdom
 \\$^3$Department of Astronomy, University of Virginia, Charlottesville, VA 22904-4325, USA
 \\$^4$Instituto de Astrofisica de Canarias, 38205 La Laguna, Tenerife, Spain
 \\$^5$Department of Astronomy, University of Texas at Austin, Austin, TX 78712, USA
 \\$^6$Department of Physics and Astronomy, Texas Christian University, Fort Worth, TX 76129, USA
 \\$^7$Departamento de Ciencias Fisicas, Universidad Andres Bello, Av. Republica 220, Santiago, Chile
 \\$^8$Institut Utinam, CNRS UMR 6213, Universit\'e de Franche-Comt\'e, OSU THETA Franche-Comt\'e-Bourgogne, Observatoire de Be\-\\san\c{c}on, BP 1615, 25010 Besan\c{c}on Cedex, France
 \\$^9$Universidad de Chile, Av. Libertador Bernardo O\'{}Higgins 1058, Santiago de Chile
 \\$^{10}$Key Laboratory for Research in Galaxies and Cosmology, 
Shanghai Astronomical Observatory, Chinese Academy of 
Sciences,\\ 80 Nandan Road Shanghai 200030, China
 \\$^{11}$Instituto de Astrofisica, Pontificia Universidad Cat\'{o}lica de Chile, Av. Vicuna Mackenna 4860, 782-0436 Macul, Santiago, Chile
 \\$^{12}$Unidad de Astronom\'{i}a, Universidad de Antofagasta, Avenida Angamos 601, Antofagasta 1270300, Chile
 \\$^{13}$Apache Point Observatory and New Mexico State University, P.O. Box 59, Sunspot, NM, 88349-0059, USA
 \\$^{14}$Sternberg Astronomical Institute, Moscow State University, Moscow
}

\date{Accepted  . Received  ; in original form 2016}

\pagerange{\pageref{firstpage}--\pageref{lastpage}} \pubyear{ }

\maketitle

\label{firstpage}

\begin{abstract}
Multiple populations revealed in globular clusters (GCs) are important windows to the formation and evolution of these stellar systems. The metal-rich GCs in the Galactic bulge are an indispensable part of this picture, but the high optical extinction in this region has prevented extensive research. In this work, we use the high resolution near-infrared (NIR) spectroscopic data from APOGEE to study the chemical abundances of NGC 6553, which is one of the most metal-rich bulge GCs. We identify ten red giants as cluster members using their positions, radial velocities, iron abundances, and NIR photometry. Our sample stars show a mean radial velocity of $-0.14\pm5.47$ km s$^{-1}$, and a mean [Fe/H] of $-0.15\pm 0.05$. We clearly separate two populations of stars in C and N in this GC for the first time. NGC 6553 is the most metal-rich GC where the multiple stellar population phenomenon is found until now. Substantial chemical variations are also found in Na, O, and Al. However, the two populations show similar Si, Ca, and iron-peak element abundances. Therefore, we infer that the CNO, NeNa, and MgAl cycles have been activated, but the MgAl cycle is too weak to show its effect on Mg. Type Ia and Type II supernovae do not seem to have significantly polluted the second generation stars.  Comparing with other GC studies, NGC 6553 shows similar chemical variations as other relatively metal-rich GCs. We also confront current GC formation theories with our results, and suggest possible avenues for improvement in the models.

\end{abstract}

\begin{keywords}
globular clusters: individual: NGC 6553 -- Galaxy: bulge  -- stars: abundances -- stars: evolution
\end{keywords}

\section{Introduction}

The longstanding notion that Galactic globular clusters (GCs) are quintessential simple stellar populations is now challenged by the discoveries of multiple populations (MPs) in an increasing number of GCs. The MP phenomenon has now been seen in main-sequence (MS), subgiant branch (SGB), red giant branch (RGB), horizontal branch (HB), and asymptotic giant branch (AGB) stars (\citealt{Meszaros2015,GH2015,Gratton2012}, and references therein). 
Among these stellar phases, giant stars are more luminous and thus more accessible for detailed studies. Color magnitude diagrams (CMDs) show prominent multiple RGBs in the cluster $\omega$ Cen when viewed in a number of different filters (e.g., \citealt{Lee1999, Pancino2000}). In particular, UV filters are the best at revealing MPs, since the UV includes a number of strong molecular bands containing especially C and N, which are prime elements affected by MPs. The first cluster abundance anomalies were reported in \citet{Osborn1971}, where two stars in M 5 and M 10 showed high N abundances in their DDO photometry. The \Stromgren $c_1$ index and the Washington C filter are also efficient in separating MPs in GCs \citep{Yong2008, Cummings2014}. Recently, the {\it Hubble Space Telescope (HST) UV Legacy Survey of Galactic GCs} \citep{Piotto2015} has observed 54 GCs through the F275W, F336W, F438W filters of the Wide Field Camera 3 (WFC3) on board $HST$ and found MPs in their entire sample with a bewildering array of detailed behavior. The first paper of that survey successfully separates at least five different populations along the MS and the RGB of the cluster NGC 2808 \citep{Milone2015}. 

While photometry is a time efficient way of revealing MPs, high resolution spectroscopy allows deeper insight into GC formation and internal stellar evolution by providing detailed elemental abundances for a number of elements with a variety of nucleosynthetic origins. N$-$C, Na$-$O, Al$-$Mg anti-correlations, and somewhat less frequently Si$-$Al correlations have been observed and discussed in numerous works: for example, \citet{Gratton2004, Carretta2009b, Carretta2009a, Carretta2010, Villanova2011, Gratton2011, Carretta2014, Carretta2015, Gratton2015}. These correlations are signatures of specific nuclear cycles. In particular, the CNO, NeNa, and MgAl modes of hydrogen burning are suggested to be responsible for the observed correlations \citep[e.g.,][]{Arnould1999, Carretta2009a, Ventura2013}.
Several possible astrophysical sites for the stars of the first generation that pollute the environment leading to a distinct second generation have been proposed, e.g., AGB stars \citep{Ventura2011, Ventura2013}, fast rotating massive stars \citep{Decressin2007}, massive binaries \citep{deMink2009}, and super-massive stars \citep{Denissenkov2014}. 

The Galactic bulge (GB) is one of the most massive and likely the oldest components of the Milky Way and its stars are direct links to the pristine formation mechanisms of the early Galaxy \citep{Schultheis2015, Howes2016}. However, despite its proximity and central role as a primordial component of the Galaxy, the GB has resisted thorough investigation due to high foreground extinction that strongly limits optical observations. Until recently, detailed spectroscopic studies with multi-object spectrographs have been mainly explored in a few low extinction windows, e.g., Baade's window and Plaut's field \citep{Zoccali2008,Hill2011,Johnson2011,Ness2013}. Viewed through these windows, the GB field stars display more than one peak in their metallicity distribution function \citep{Babusiaux2010, Ness2013}.
However, observing in the NIR helps to minimize the generally strong extinction and opens the entire bulge for study. Therefore, a NIR high resolution multi-object spectrograph, such as that used by the Apache Point Observatory Galactic Evolution Experiment (APOGEE; \citealt{Majewski2015}), provides new opportunities to push forward our knowledge about the chemical evolution of the bulge \citep[e.g.,][]{Schiavon2016}.

GB GCs provide important insight into galaxy formation as well as its subsequent dynamical and chemical evolution \citep{Mauro2012, Bica2015, Cohen2016}. 
NGC 6553 ($\alpha_{\rm J2000} =  \rm 18^h09^m17.6^s$, $\delta_{\rm J2000} = -25^{\circ}54'31''$, $l=5.25^{\circ}$, $b = -3.03^{\circ}$) is one of the most metal-rich bulge GCs, and has Galactic coordinates in a Sun-centered system of X = +5.9 kpc, Y = +0.5 kpc, and Z = $-$0.3 kpc\footnote{http://physwww.physics.mcmaster.ca/$\sim$harris/mwgc.dat}. Therefore, NGC 6553 is at the near edge of the bulge and, accounting for a reddening ($E_{B-V}$) of $0.63$, has a luminosity ($M_V$) of $-7.77$ \citep{Harris1996}.  If $M_V$ is used as a rough proxy of GC mass, NGC 6553 is an intermediate-mass example in the bulge GC mass distribution. Note that we see a spread in the current values of Galactic coordinates, reddening, and luminosities from \citet{Guarnieri1998}, \citet{Harris1996}, and \citet{Valenti2010}\footnote{http://www.bo.astro.it/$\sim$GC/ir\_archive/Tab1\_new.html},  so these values may not be known to better than $\sim10-20$\%. 
In terms of GC metallicity, \citet{AlvesBrito2006} noticed a significant range in literature values,  $-0.55 <$ [Fe/H] $< -0.06$ (a complete list of literature results can be found in the next section).  
\citet{Johnson2014} measured the Na, Mg, Al, Si, Ca, Cr, Fe, Co, Ni, and Cu abundances in twelve cluster members, and found the chemical pattern of NGC 6553 agrees with other bulge field stars, except for larger Na scatter. They suggested that this scatter may be caused by additional self-enrichment.

Thanks to the lower extinction and more available CNO molecular lines in the APOGEE survey, we recently found substantial chemical variations in several bulge GCs, e.g., NGC 6553, NGC 6528, and Terzan 5 (Schiavon et al. 2016, submitted). In this paper, we present a more detailed analysis of the chemical pattern observed in the giant stars of NGC 6553 ($\S$\ref{sect:data}). We clearly separate two groups of stars in C and N ($\S$\ref{sect:tg}). We investigate the APOGEE chemical abundances, and supplement it with two recent high resolution spectroscopic studies ($\S$\ref{sect:re}). We compare our results to the general picture outlined by literature GC studies, and discuss the possibility of applying AGB polluting models to explain the chemical pattern in NGC 6553 ($\S$\ref{sect:dis}). Finally, a brief summary of the results and a look to the future are given in $\S$\ref{sect:con}.

\begin{figure}
\centering
\includegraphics [width=0.45\textwidth]{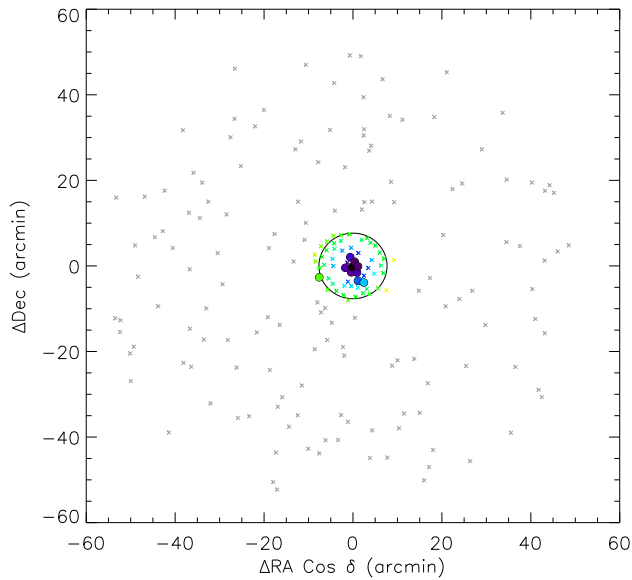} 
\caption{Locations of the stars in the 7 deg$^2$ APOGEE field of NGC 6553. The solid circle indicates cluster tidal radius ($R_{t}$). Non-members outside 1.5$R_{t}$ are the grey crosses. Stars inside 1.5$R_{t}$ are color-coded, according to the color bar shown in Figure \ref{fig:VVV1}. Cluster members and non-members are labeled by filled  circles and crosses, respectively.}\label{fig:VVV2}
\end{figure}

\begin{figure}
\centering
\includegraphics [width=0.45\textwidth]{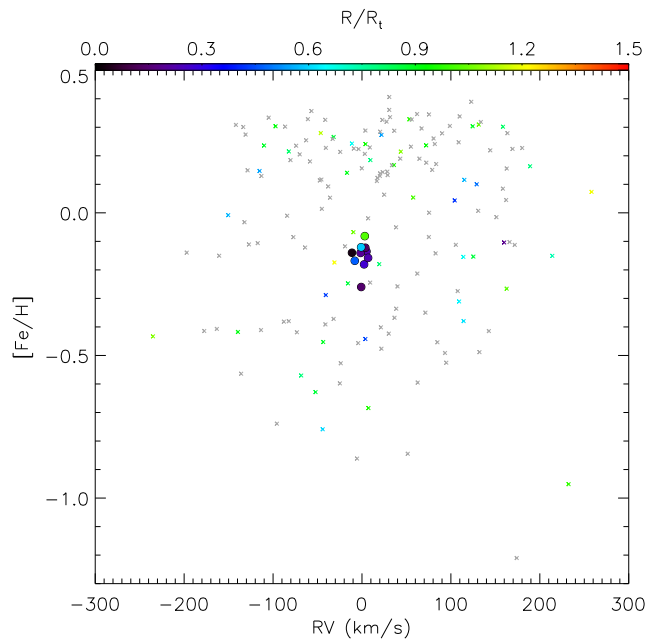} 
\caption{[Fe/H] versus radial velocity. Symbol meanings are the same as in Figure \ref{fig:VVV2}. }\label{fig:VVV1}
\end{figure}

\section{Sample Selection and Data Reduction}
\label{sect:data}

APOGEE \citep{Majewski2015} was one of the programs operating during the Sloan Digital Sky Survey III (SDSS-III, \citealt{Eisenstein2011}). The multi-object NIR fiber spectrograph on the 2.5 m telescope at Apache Point Observatory \citep{Gunn2006} delivers high-resolution ($R\sim$22,500) $H$-band spectra ($\lambda = 1.51 - 1.69$ $\mu$m), and the APOGEE survey targeted a color-selected sample that predominately consists of good stars across the Milky Way.  APOGEE data reduction software is applied to reduce multiple 3D raw data cubes into calibrated, well-sampled, combined 1D spectra \citep{Nidever2015}. In addition, the APOGEE Stellar Parameter and Chemical Abundances Pipeline (ASPCAP; \citealt{GP2016}) derives stellar parameters and elemental abundances by comparing observed spectra to libraries of theoretical spectra \citep{Shetrone2015, Zamora2015} to find the closest model match, using $\chi^2$ minimization in a multidimensional parameter space. Through SDSS Data Release 12 (DR12), up to 15 chemical elements were identified, and measured abundances provided. The calibrations of stellar parameters and abundances from the pipeline were described in \citet{Holtzman2015}. They compared the APOGEE abundances with those from other high resolution spectroscopic studies, and suggested that the internal scatter of the calibrated abundances is generally between 0.05 and 0.09 dex, while the external accuracy may be good to 0.1$-$0.2 dex.
A new data release (SDSS DR13) is now publicly available \citep[Holtzman et al. 2016, in prep.;][]{SDSS2016}. DR13 is a re-release of all APOGEE-1 data between May 2011 and July 2014, where pipeline and calibration improvements made since DR12 are integrated, and more element species are identified, i.e., \ci, \tiii, P, Cr, Co, Cu, Ge, and Rb. Therefore, we decided to use the results from DR13. Note that all the element abundances are derived under LTE (Local Thermodynamic Equilibrium) assumption. Non-LTE effects for the NIR lines are poorly known but currently being investigated \citep[e.g.,][]{Bergemann2013,Bergemann2015}.

\begin{figure*}
\centering
\includegraphics [width=0.8\textwidth]{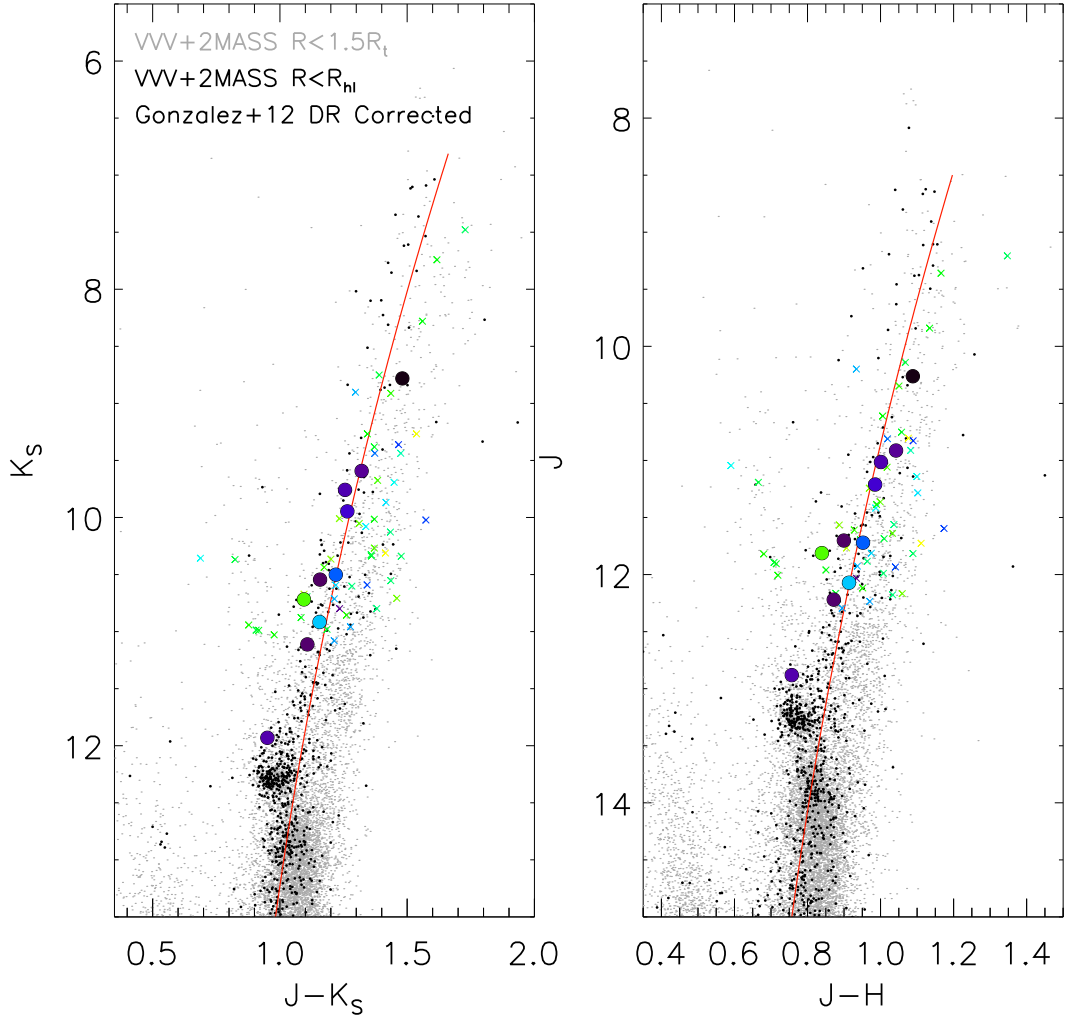} 
\caption{Differential reddening-corrected CMD of NGC 6553 from PSF photometry of VVV imaging, supplemented with bright stars from 2MASS. The cluster fiducial sequence is indicated as a red solid line. The grey dots are stars within 1.5$R_{t}$ (tidal radius), and the black dots are stars within the half-light radius.  APOGEE targets inside 1.5$R_{t}$ are color-coded as Figure \ref{fig:VVV2}. Cluster members and non-members determined in this paper are labeled by filled circles and crosses, respectively. See text for more details. }\label{fig:VVV4}
\end{figure*}

To gain as much information on cluster membership as possible, our spectroscopic targets were positionally matched to near-IR PSF photometry based on imaging from the \textit{VISTA Variables in the Via Lactea} (VVV) survey \citep{Minniti2010}.  The resulting $JHK_{S}$ photometric catalogs have photometric zeropoint uncertainties and an astrometric rms of $\sim$0.02 mag and 0.2$\arcsec$, respectively, with respect to the Two Micron All Sky Survey Point Source Catalog \citep[2MASS PSC,][]{Skrutskie2006}. Examples and details regarding 2MASS-calibrated PSF photometry of preprocessed VVV imaging are presented elsewhere \citep{Chene2012,Mauro2012,Mauro2013,Cohen2014}.  In addition, the VVV PSF catalogs have been merged with bright stars from the 2MASS PSC that are saturated in the VVV images.  The near-IR photometric properties and CMD of NGC 6553 are discussed in detail in \citet{Cohen2016}.  In addition to employing the raw 2MASS-calibrated near-IR CMDs, we have also employed the reddening map of \citet{Gonzalez2012} to correct for spatially variable reddening, but we apply this map in a differential sense relative to the value of $E(J-K_{S})$ that they report at the center of NGC 6553.  Although reddening maps are available at higher spatial resolution close to the cluster center \citep{AlonsoGarcia2012}, these maps rely on the use of cluster members, and therefore do not extend over the full tidal radius of the cluster.

We select candidate cluster members by leveraging together all of the information at hand, including positions, radial velocities and abundances from APOGEE, and the aforementioned PSF photometry.  We begin by considering only stars within 1.5 times the cluster tidal radius ($R_{t}$) from the \citealt{Harris1996} catalog.  We include stars slightly beyond $R_{t}$ because the tidal radius resulting from King profile fits, especially when relying on optical integrated light profiles, is not always well constrained \citep[e.g.,][]{McLaughlin2005,Cohen2014}. Extratidal stars identified via abundances consistent with their host clusters have been identified spectroscopically in some cases \citep[e.g.,][]{Kunder2014}.  Although this topic is beyond the scope of the present investigation, there were no stars in our sample falling slightly outside our radial cut that had photometry, velocities and metallicities clearly indicative of membership in NGC 6553.  In Figure~\ref{fig:VVV2}, we plot the spatial location of all stars in the sample (within the 7 deg$^2$ APOGEE field), color-coded by radius from the center of NGC 6553, with stars having $R>1.5R_{t}$ shown in grey.  Stars that we consider to be members are plotted in Figure~\ref{fig:VVV2} as filled circles, and non-members are shown using small crosses; the large circle indicates the cluster tidal radius.   Figure~\ref{fig:VVV1} shows these same stars in the plane of [Fe/H] versus radial velocity, both derived from APOGEE, and we see a clear concentration of stars that are spatially close to the center of NGC 6553  and having properties in excellent agreement with literature values for the cluster [Fe/H] and radial velocity (see the last paragraph of this section for a complete list of literature results).  At this point, we have a preliminary cluster candidate list that is quite conservative, with all excluded stars having radial velocities differing from the cluster mean by $>$50 km/s (compared to typical GC central velocity dispersions of $<$20 km/s; \citealt{Harris1996,McLaughlin2005,Watkins2015}), or [Fe/H] differing from the cluster mean by $>$0.2 dex (compared to the $<$0.1 dex relative precision of APOGEE abundances; see Table \ref{tab1}), or in most cases, both.  As a final check on cluster membership, we compare the locations of candidate members in the differential reddening-corrected near-IR CMDs against the expected locus of NGC 6553 in Figure~\ref{fig:VVV4}.  There, we plot stars inside the cluster half-light radius in black and those within 1.5$R_{t}$ in grey to highlight the relative locations of the cluster and field (e.g.,~Galactic bulge and disk) in the CMD.  In addition, an empirical cluster red giant branch fiducial sequence constructed from statistically decontaminated photometry \citep{Cohen2016} is shown as a red line.  Given the differential reddening towards this cluster, as well as the formal uncertainties of the \citet{Gonzalez2012} differential reddening corrections, we do not make rigorous CMD cuts, but rather simply exclude all candidate members passing the aforementioned spatial, metallicity and velocity cuts that have CMD locations highly discrepant with the locus of the NGC 6553 red giant branch.  Specifically, four candidates were considered non-members because they have $(J-K_{S})$ colors more than 0.1 mag from the cluster fiducial sequence.  An additional star with a radial velocity and CMD location consistent with cluster membership was excluded because it does not have reliable abundances from APOGEE. We note that two of the stars in the NGC 6553 sample selected by Schiavon et al. (2016, submitted) are excluded in the present work due to their CMD locations.

\begin{table*}
\caption{Basic parameters for cluster members of NGC 6553.}              
\label{tab1}      
\setlength{\tabcolsep}{9pt} 
\begin{tabular}{c c c c r r r}         
\hline\hline                        
\#& APOGEE ID&RA & DEC  &  $J^{a}$  & $H^{a}$ & $K_{S}^{a}$  \\ 
&         & (deg) & (deg) & (mag) & (mag) & (mag) \\
\hline                                   
1 & 2M18084368-2557107 & 272.182001 & -25.952997 & 11.812 & 10.973 & 10.718 \\
2 & 2M18090968-2554574 & 272.290370 & -25.915968 & 12.881 & 12.124 & 11.930 \\
3 & 2M18091466-2552275 & 272.311107 & -25.874331 & 11.211 & 10.226 & 9.946 \\
4 & 2M18091564-2556008 & 272.315203 & -25.933556 & 10.913 & 9.871 & 9.592 \\
5 & 2M18091666-2554424 & 272.319437 & -25.911798 & 10.262 & 9.174 & 8.781 \\
6 & 2M18091912-2553326 & 272.329703 & -25.892410 & 11.702 & 10.802 & 10.544 \\
7 & 2M18092147-2556039 & 272.339462 & -25.934441 & 11.014 & 10.013 & 9.758 \\
8 & 2M18092234-2554381 & 272.343108 & -25.910591 & 12.220 & 11.348 & 11.112 \\
9 & 2M18092241-2557595 & 272.343397 & -25.966530 & 11.721 & 10.769 & 10.501 \\
10 & 2M18092826-2558233 & 272.367760 & -25.973152 & 12.071 & 11.158 & 10.916 \\
\end{tabular}
\setlength{\tabcolsep}{4pt}
\begin{tabular}{c r c c c c c c c r c l} 
\hline
\#& RV& $\delta_{\rm RV}^{b}$& [Fe/H]& $\delta_{\rm [Fe/H]}^{b}$& T$_{\rm eff}$ & $\delta_{\rm T_{eff}}^{b}$ & $\log{\rm g}$& $\delta_{\log{\rm g}}^{b}$ &  SNR &{\footnotesize PERSIST}$^{c}$& WARN$^{d}$\\ 
& \multicolumn{2}{c}{(km s$^{-1}$)} &\multicolumn{2}{c}{(dex)} &\multicolumn{2}{c}{(K)} &\multicolumn{2}{c}{(dex)}  &  &  & \\
\hline 
1  &  3.17 &  0.02 & -0.08 &  0.03 & 4176.6 &  69.3 &  1.76 &  0.08 & 92 &  HIGH &  \\
2  &  5.08 &  0.05 & -0.14 &  0.03 & 4716.0 &  69.3 &  2.38 &  0.08 & 46 &  MED & N, SN \\
3  &  2.39 &  0.01 & -0.18 &  0.03 & 4069.0 &  69.3 &  1.62 &  0.08 & 138 &  & N  \\
4  & -1.54 &  0.01 & -0.14 &  0.03 & 3971.7 &  69.3 &  1.45 &  0.08 & 181 &  MED & N  \\
5  & -11.26 &  0.00 & -0.14 &  0.03 & 3811.7 &  69.3 &  1.23 &  0.08 & 261 &  MED & N  \\
6  &  3.79 &  0.02 & -0.12 &  0.03 & 4357.9 &  69.3 &  2.06 &  0.08 & 88 &  MED & N  \\
7  &  6.87 &  0.01 & -0.16 &  0.03 & 4049.3 &  69.3 &  1.52 &  0.08 & 153 &  MED & N  \\
8  & -0.88 &  0.02 & -0.26 &  0.03 & 4345.8 &  69.3 &  2.11 &  0.08 & 82 &  MED &  \\
9  & -8.12 &  0.01 & -0.17 &  0.03 & 4047.8 &  69.3 &  1.43 &  0.08 & 110 &  HIGH &  \\
10  & -0.94 &  0.02 & -0.12 &  0.03 & 4345.1 &  69.3 &  2.01 &  0.08 & 89 &  HIGH & N  \\
\hline                                             
\end{tabular}

\raggedright{$^a$ Differential reddening corrected magnitude.}\\
\raggedright{$^b$ Measurement error.}\\
\raggedright{$^c$ PERSIST: spectrum has a significant number ($>$20\%) of pixels in the high (or medium) persistence region.}\\
\raggedright{$^d$ N\_WARN: parameter value is within $1/2$ grid spacing of the synthetic spectrum grid edge for nitrogen. SN\_WARN: SNR$<$70.}
\end{table*}

From the coordinate information, these ten cluster members have no overlap with the spectroscopic studies of NGC 6553 published after the year 2006, which include \citet{AlvesBrito2006}, \citet{Zoccali2008}, \citet{Gonzalez2011}, \citet{Johnson2014}, and \citet{Dias2015}. 
Most of the APOGEE spectra for our sample have signal to noise ratio (SNR) higher than 80, except the second star in Table \ref{tab1}. Each star in our sample has a single one hour visit, so the SNR is correlated with the stellar brightness. The SNR plays an important role in estimating the uncertainties of element abundances \citep{Majewski2015}.  Three stars are labeled as ``PERSIST\_HIGH'' and six are labeled as ``PERSIST\_MED''. Persistence,  where a significant fraction of accumulated charge is released over a long period of time, is particularly strong in one of the detectors (1.51$-$1.58 $\mu$m) used in the SDSS III/APOGEE survey \citep{Nidever2015}. However, by comparing the results from spectra containing persistence pixels versus those without, \citet{Holtzman2015} suggests that persistence does not dramatically impact the parameters in DR12 data. 
 
From APOGEE measurements, our ten cluster members show a mean RV of $-0.14\pm5.47$\footnote{Standard deviation.} km s$^{-1}$. Our mean RV agrees with the other determinations:  $4\pm7.1$ km s$^{-1}$ \citep{Cohen1999}, $1.6\pm6$ km s$^{-1}$ \citep{Melendez2003}, $-1.86\pm2.01$ km s$^{-1}$ \citep{AlvesBrito2006}, $-2.03\pm4.85$ km s$^{-1}$ \citep{Johnson2014}, and $6\pm8$ km s$^{-1}$ \citep{Dias2015}.
In addition, the APOGEE data yield a mean [Fe/H] of $-0.15\pm 0.05$. This GC has diverse metallicity results from the literature: $-0.55\pm0.2$ \citep{Barbuy1999}, $-0.16\pm0.08$ \citep{Cohen1999}, $-0.7\pm0.3$ \citep{Coelho2001}, $-0.3\pm0.2$ \citep{Origlia2002}, $-0.2\pm0.1$ \citep{Melendez2003}, $-0.2 \pm0.02$ \citep{AlvesBrito2006}, $-0.11\pm0.07$ \citep{Johnson2014}, and $-0.13\pm0.02$ \citep{Dias2015}. We notice that as more high resolution and high S/N spectra have become available in recent years, the cluster metallicity is converging to between $-0.1$ and $-0.2$ dex, which agrees well with our result from APOGEE.

\section{Two Populations of Stars with Distinct Chemical Abundances}
\label{sect:tg}

\begin{figure*}
\centering
\includegraphics [width=0.9\textwidth]{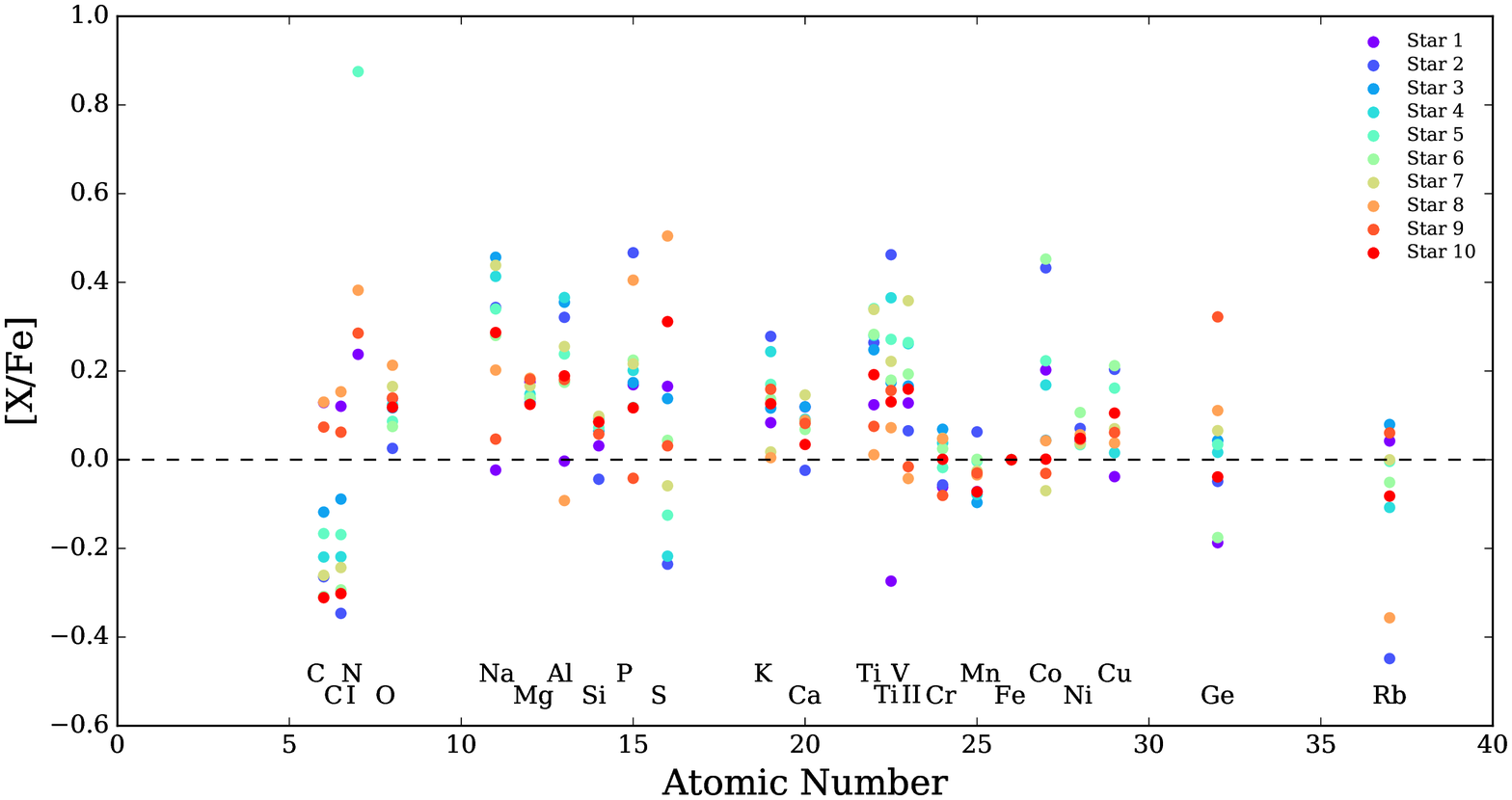} 
\caption{Calibrated element abundances from APOGEE measurements as a function of atomic number for NGC 6553 stars. Cluster members are labeled with different colors. The IDs of the elements are shown at the bottom, where \ci~and \tiii~are offset by 0.5 atomic number for clarity. Note that six stars with the ``N\_WARN'' flag (except star 5; Table \ref{tab1}) have no calibrated N abundances.}\label{fig:an6553}
\end{figure*}

Figure \ref{fig:an6553} shows the calibrated chemical abundances from ASPCAP as a function of atomic number for NGC 6553 stars; see Table \ref{tab2} for detailed information. In DR13, [C/Fe] is calculated using the molecular bands (mainly CO), and [\ci/Fe]~measures the C abundance from atomic lines. In the case of Ti, [Ti/Fe] is based only on \tii~lines, and [\tiii/Fe] is from a single \tiii~line. 

Abundances of the light elements involved in proton capture processes (C, N, O, F, Na, Mg, Al, and Si) are known to vary in GCs, which is the main chemical evidence for MPs \citep{Kraft1979,Kraft1994,Carretta2009b,Carretta2009a}. But the scarcity of strong C and N lines in the optical increases the difficulty of MP studies. Moreover, the C and N molecular line regions in the near-UV can be too crowded for metal-rich stars, which increases the difficulty of line identification \citep{Boberg2016}. 
This situation is eased in the NIR, where more C, N, and O lines are available and these lines are less crowded.

\begin{figure}
\centering
\includegraphics [width=0.5\textwidth]{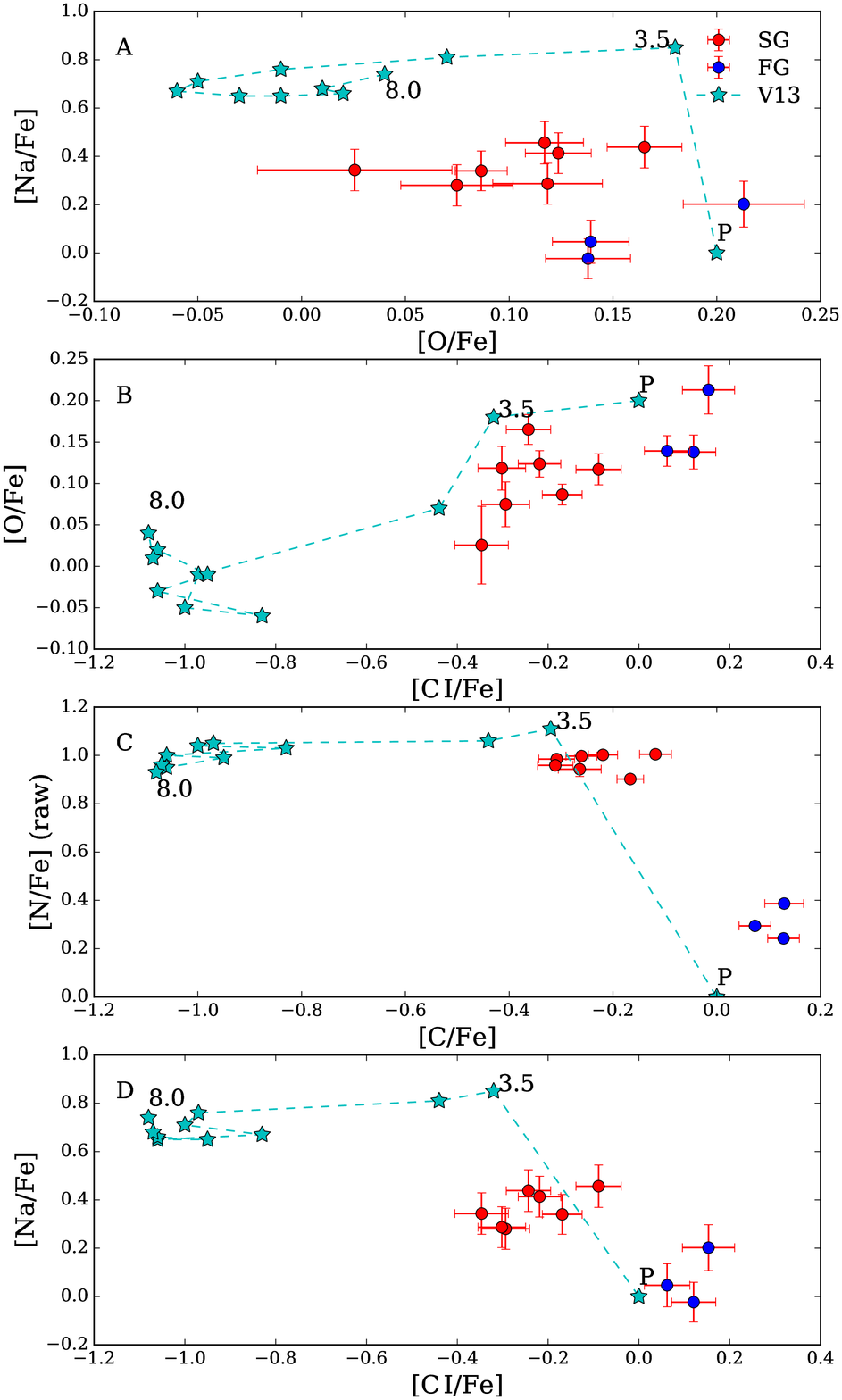} 
\caption{Two generations of stars in the parameter space of [C/Fe], [\ci/Fe], [N/Fe] (raw), [O/Fe], and [Na/Fe]. The presumed first generation stars are labeled as blue circles, and the second generation stars are labeled as red circles. The error bars indicate the measurement errors. The cyan stars are pure yields from the metal-rich AGB models of \citet{Ventura2013}. The primordial abundances are labeled with ``P'', and the initial mass of the stars in solar mass units are indicated by numbers. See text for more details about the models.}\label{fig:CNONa}
\end{figure}

\begin{figure}
\centering
\includegraphics [width=0.5\textwidth]{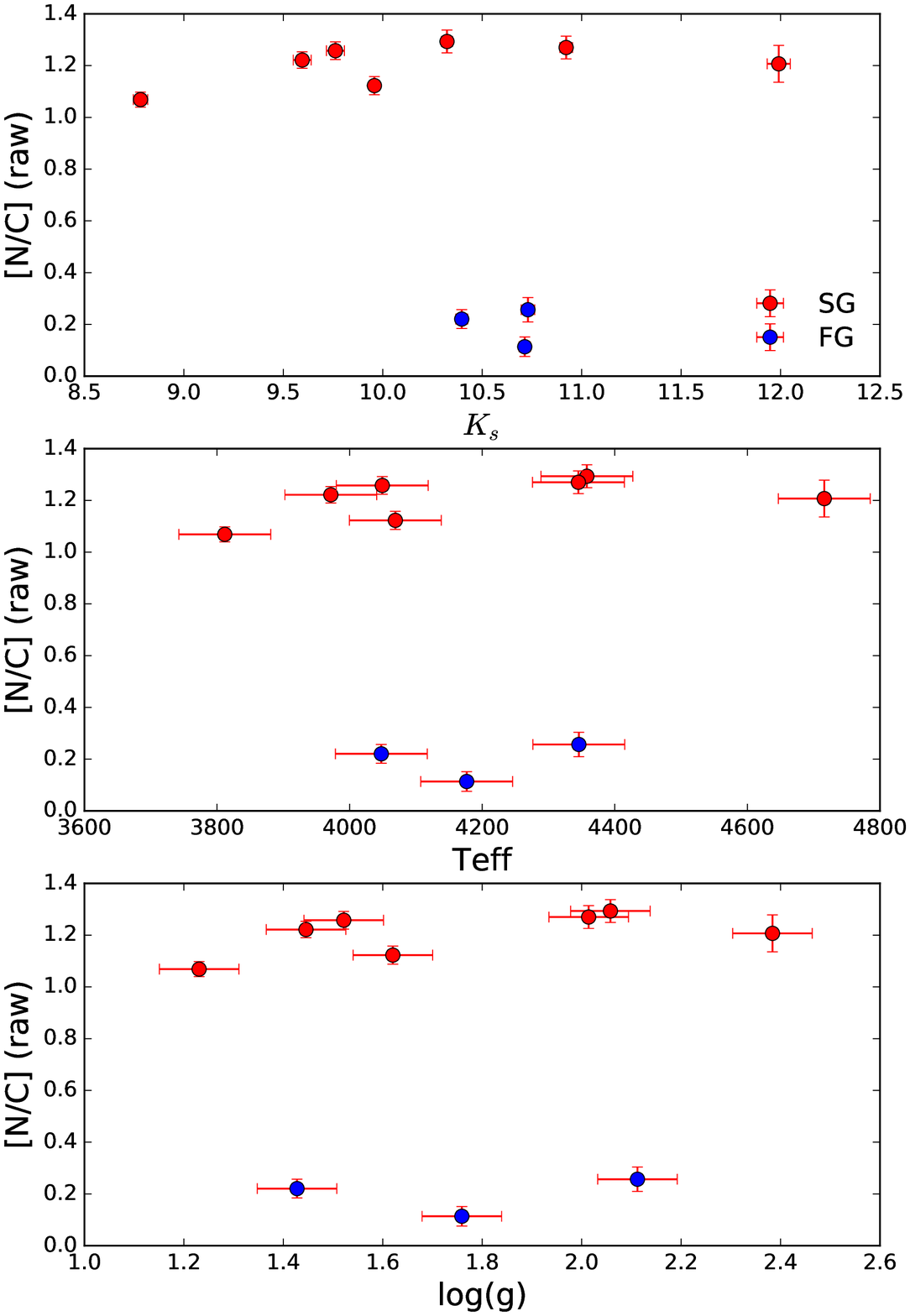} 
\caption{[N/C] (raw) as a function of $K_s$, T$_{\rm eff}$ and $\log{\rm g}$. The first generation stars are labeled as blue circles, while the second generation stars are labeled as red circles.}\label{fig:CNTeff}
\end{figure}

In our APOGEE NGC 6553 sample, seven stars have an ``N\_WARN'' flag (Table \ref{tab1}), indicating that the nitrogen abundance is fitted within a $1/2$ grid spacing of the synthetic spectrum grid edge. A quick examination of the chemical pattern of each star in Figure \ref{fig:an6553} reveals two possible groups of stars with distinct calibrated C (as measured from molecular lines), \ci~(as measured from atomic lines), and N\footnote{Six stars with the ``N\_WARN'' flag (except Star 5) have no calibrated N abundances assigned in DR13.} abundances. To further explore this idea, we plot the patterns of [Na/Fe]$-$[O/Fe], [O/Fe]$-$[\ci/Fe], [N/Fe] (raw)$-$[C/Fe], and [Na/Fe]$-$[\ci/Fe] in Figure \ref{fig:CNONa}.  [N/Fe] (raw) abundances are converted from [N/M]. We use ``raw'' here to differentiate from calibrated abundances. The [N/M] values come from a global fit to the spectrum, which is done simultaneously with the main stellar parameters, e.g., T$_{\rm eff}$, log(g), and etc. The stars with the ``N\_WARN'' flag may have real [N/Fe] abundances that are higher than [N/Fe] (raw). Therefore the [N/Fe] (raw) abundances should be used with extra caution. An independent confirmation of conclusions drawn solely from [N/Fe] (raw) would be helpful. Panels C and D of Figure \ref{fig:CNONa} clearly show two groups of stars with distinct chemical abundances of [C/Fe], [\ci/Fe], and [N/Fe] (raw). The best-fit [Na/Fe] abundances are different in the two groups of stars, but if measurement errors are considered, a more continuous variation in [Na/Fe] is possible. [O/Fe] abundances of the two groups of stars also have overlap.
A very similar situation was also found in M4 \citep{Villanova2011}. Interestingly, \citet{Meszaros2015} suggested that the separation between N-weak and N-strong groups increases with cluster metallicity. Our confirmation of C and N bimodality in the metal-rich GC NGC 6553 agrees with this hypothesis.

The chemical variations in GCs may have several possible explanations. If chemical variations are caused by the first dredge-up and extra mixing in the RGB phase \citep{Boothroyd1999, Charbonnel2007}, then the associated theories predict that [N/C] should be an increasing function of absolute magnitude or stellar mass. Using the $K_s$ apparent magnitude as a proxy\footnote{Assuming similar distances for cluster members} for intrinsic luminosity, however, we find no such correlation in NGC 6553 (top panel of Figure \ref{fig:CNTeff}).  \citet{Meszaros2015} mention that ASPCAP temperatures in metal-poor GCs may be subjected to possible offsets with respect to ones based on photometry. NGC 6553 is one of the most metal-rich GCs, therefore the temperature offset should not be severe here. Furthermore, while there may be concern that the chemical variations are caused by temperature or surface gravity effects,  the [N/C] vs. T$_{\rm eff}$ and [N/C] vs. log(g) plots show this concern is not valid, because the two groups of stars have similar temperatures and surface gravities (middle and bottom panels of Figure \ref{fig:CNTeff}). Thus, the self-enriched two stellar generation scenario seems to be a viable theory to explain the chemical variations that we observed. Following the literature convention, we name these two stellar generations as First Generation (FG), and Second Generation (SG). The FG in NGC 6553 includes stars 1, 8, and 9 (Table \ref{tab1}), while the rest of the stars are SG. The FG stars represent almost 1/3 of our sample (see $\S$\ref{sect:liter} for more discussion). 



\section{Chemical Abundances of NGC 6553}
\label{sect:re}

Before we further investigate the other element abundances of NGC 6553 derived by APOGEE, it is helpful to know more about the error budgets of APOGEE measurements. \citet{Holtzman2015} estimated that the internal scatter for typical APOGEE abundance values is between 0.05 and 0.09 dex, while the external scatter is about 0.1$-$0.2 dex.
 Therefore, internally comparing the APOGEE calibrated abundances is more reliable. 
Recently, \citet{Souto2016} manually derived the chemical abundances for 12 stars in NGC 2420, one of the calibrating clusters for ASPCAP, and compared the results with the DR13 calibrated abundances. The derived mean metallicity for NGC 2420 stars is $-0.16$, which is very close to the mean metallicity of NGC 6553 stars. Therefore the Souto et al. study is informative for understanding the calibrated ASPCAP abundances that we use in this paper. Souto et al. found generally good agreements (i.e., $\leqslant$0.1 dex) in the chemical abundances between the manually derived results and the DR13 calibrated results, except for the elements Na, Al, and V. Note that the difference between results derived manually and that from DR13 is an indication of the external abundance scatter.

Nevertheless, we also plot the results from two recent high resolution studies of NGC 6553 in our figures: \citet[][AB06]{AlvesBrito2006} and \citet[][J14]{Johnson2014}. We include these two representative samples because the former sample was observed by UVES, which has very high spectral resolution (R$=$47,000) in the optical, and the latter one has decent sample size (N$=$12), though at lower spectral resolution (R$=$20,000). 
We use the measurements from J14 here, instead of the measurements from \citet{Zoccali2008} and \citet{Gonzalez2011}, for the following reasons:
(1) the J14 and the Gonzalez et al. samples are selected from the Zoccali et al. sample. We pick one sample to avoid duplication;
(2) the twelve cluster members in the J14 sample are identified by the [Fe/H] vs. RV diagram;
(3) the results from J14 and Gonzalez et al. samples agree reasonably well, and J14 presents more abundance measurements.
Unfortunately, we have no stars in common with the above samples.

\subsection{The CNO, NeNa and MgAl cycles}
Na and Al originally are synthesised by C and Ne burning in massive stars \citep{Arnett1971, Clayton2007, Woosley1995}, but they can also be produced through the NeNa and MgAl cycles during the hydrogen burning \citep{Arnould1999}. The noble gas Ne cannot be detected in cool stars. Meanwhile, O, as part of the CNO cycle, shows a decreasing trend with core temperature \citep{Arnould1999}. Contrary to  C and N, which are almost unmeasurable from optical spectra, O has a few forbidden lines in this optical region, e.g., [\oi] at 6300 and 6363 \AA. Though the exact nature of the polluters responsible for O and Na variations are still under study (see $\S$\ref{sect:dis}), the Na$-$O anti-correlation is broadly observed in GCs \citep[e.g.,][]{Sneden2000}, and has been used to distinguish the two generations of stars \citep[e.g.,][]{Carretta2009a}. It is also seen in this work (Panel A of Figure \ref{fig:CNONa}). At the same time, the O$-$\ci~correlation presented in Panel B of Figure \ref{fig:CNONa} also suggests the CNO cycle is activated. 

\begin{figure}
\centering
\includegraphics [width=0.5\textwidth]{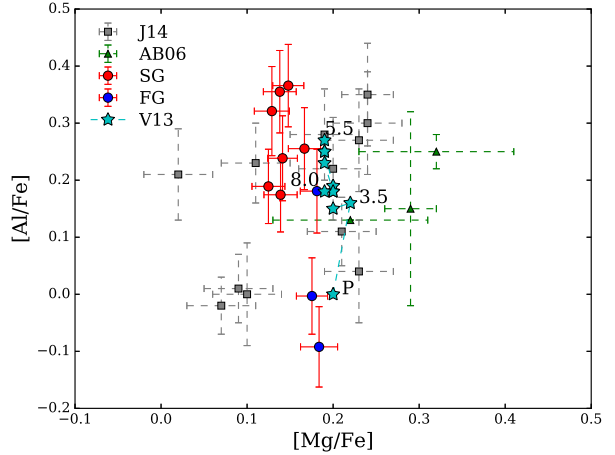} 
\caption{[Al/Fe] versus [Mg/Fe]. We also plot the measurements and uncertainties from \citet{Johnson2014} (grey squares) and \citet{AlvesBrito2006} (green triangles). The rest of the symbols are explained in Figure \ref{fig:CNONa}. }\label{fig:AlMg}
\end{figure}

The Al$-$Mg anti-correlation is suggestive of the MgAl cycle. \citet{Carretta2009b} found that the Al$-$Mg anti-correlation is not always present in their sample of 17 Galactic GCs. The existence of this anti-correlation may be related to the mass and metallicity of the GC. Theoretical studies \citep{Arnould1999, Ventura2013} indicate that the MgAl cycle requires higher temperature than the NeNa cycle, and the fraction of Mg that is transfered to Al is also smaller.
In Figure \ref{fig:AlMg}, the [Al/Fe] dynamic range reaches $\sim0.5$ dex, and FG stars generally show smaller [Al/Fe] than SG stars. Note that one FG star has [Al/Fe] comparable to the lowest [Al/Fe] found in SG stars. This is possibly caused by the internal error on Al measurements. On the other hand, the [Mg/Fe] abundances show a much smaller variation ($\sim 0.08$ dex). We see a hint of slightly larger [Mg/Fe] in FG stars. However, due to the internal scatter of the calibrated abundances from ASPCAP (0.05$-$0.09 dex), we cannot tell if the [Mg/Fe] abundances are truly different in the two generations. 
This is consistent with the studies of \citet{Carretta2009b}, who found that the Al$-$Mg anti-correlations are less prominent in metal-rich GCs, because the core temperature of the polluting stars may not be high enough to covert most Mg to Al. However, AB06 and J14 measurements show a large Mg scatter, possibly caused by the larger errors or different stars. In any case, no Al$-$Mg anti-correlation can be found in any of these samples. Clearly, the calibrated Mg abundances from ASPCAP have a smaller scatter than the optical Mg abundances, a finding that supports the notion that the former [Mg/Fe] measurements have higher quality.

\begin{figure}
\centering
\includegraphics [width=0.5\textwidth]{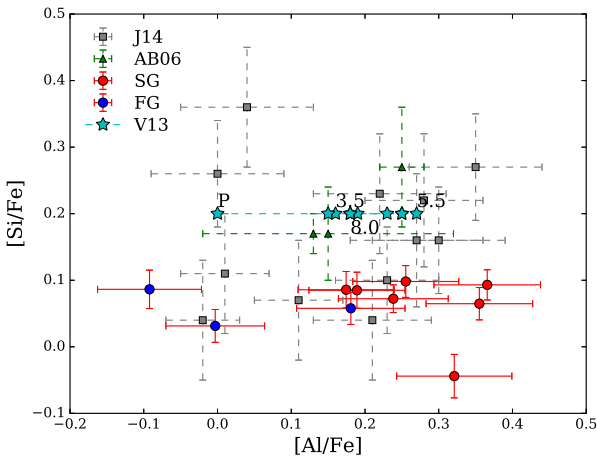} 
\caption{[Si/Fe] versus [Al/Fe]. Symbols are explained in Figures \ref{fig:CNONa} and \ref{fig:AlMg}. }\label{fig:SiAl}
\end{figure}

Silicon is suggested to be a possible ``leakage'' from the MgAl cycle when the temperature is high enough \citep{Arnould1999, Yong2005, Ventura2013}. The temperature dependence of Maxwellian-averaged reaction rates for proton captures can be used to predict when the $^{28}$Si leakage becomes predominant in the MgAl cycle (Figure 8 of \citealt{Arnould1999}).  
Note that this temperature may be metallicity- and model-dependent. 
\citet{Carretta2009b} explained their observation of lack of a Si$-$Al correlation in metal-rich GCs with the Si leakage theory. The Si$-$Al correlation is also nonexistent in the APOGEE sample of NGC 6553 stars (Figure \ref{fig:SiAl}). The Si abundances from AB06 and J14 seem to be systematically larger than the APOGEE results. This may be caused by the Si zeropoint offset issue in APOGEE pipeline and calibration \citep{Holtzman2015}. However, similar to Mg, the calibrated Si abundances from ASPCAP also show smaller scatter than the optical Si abundances, indicating that at least their internal errors are small.

\begin{figure}
\centering
\includegraphics [width=0.5\textwidth]{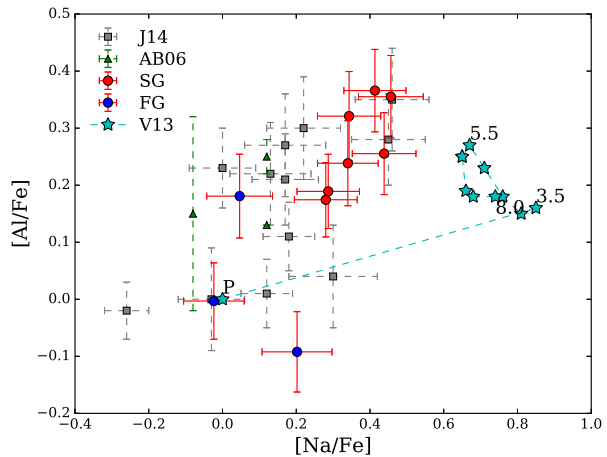} 
\caption{[Al/Fe] versus [Na/Fe]. Symbols are explained in Figures \ref{fig:CNONa} and \ref{fig:AlMg}. }\label{fig:AlNa}
\end{figure}

\begin{figure}
\centering
\includegraphics [width=0.5\textwidth]{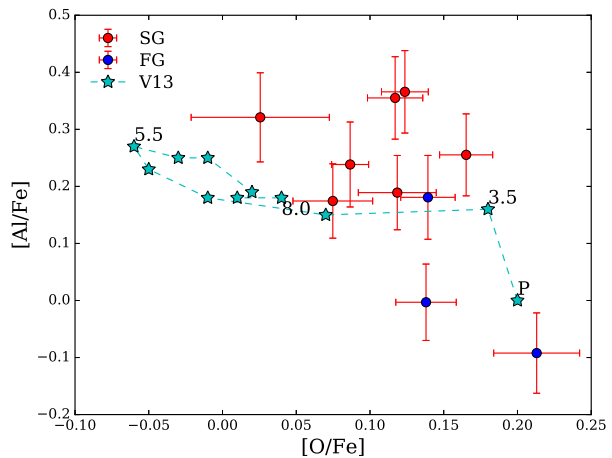} 
\caption{[Al/Fe] versus [O/Fe]. Symbols are explained in Figures \ref{fig:CNONa} and \ref{fig:AlMg}. }\label{fig:AlO}
\end{figure}

In Figure \ref{fig:AlNa}, Na positively correlates with Al for our sample, i.e., the three FG stars in general show lower Al abundances. Though the Al$-$Na correlation is not suggested in the work of J14, combining our sample, AB06 and J14 strengthens the visibility of the correlation. In general, we find larger scatter in the lower end of Na and Al abundances.  If one considers the nucleosynthetic process associated with Na, the correlation between Na and Al may suggest the MgAl cycle has begun in NGC 6553 at a low level. 
Since Al is about an order of magnitude less abundant than Mg in the Sun and even more in the primordial stars in GCs \citep{Carretta2009b}, the MgAl cycle may significantly change Al, but not Mg, as observed.
The Al$-$O anti-correlation is also suggested in other GCs (e.g., \citealt{Shetrone1996, Kraft1997, Sneden1997, Carretta2009b}, but also see \citealt{Origlia2011}). Figure \ref{fig:AlO} clearly shows that the Al$-$O anti-correlation is present in our sample.

\subsection{$\alpha$ elements}
\label{sect:alpha}

According to the nucleosynthetic processes that are associated with different $\alpha$ elements during Type II supernova (SNe), O and Mg are commonly classified as hydrostatic $\alpha$ elements, and Si, Ca, and Ti are classified as explosive $\alpha$ elements \citep{Woosley1995}. O and Mg are two of the primary $\alpha$ elements produced, and they are produced in almost the same ratio for stars of disparate mass and progenitor heavy element abundance. On the other hand, two of the heaviest explosive $\alpha$ elements, Ca and Ti, follow O and Mg in the Galactic environment (e.g., Milky Way bulge), but seem to have a substantial contribution besides Type II SNe in extreme extragalactic environment \citep[e.g., massive elliptical galaxies;][]{Worthey2014,TangB2014}. 

\begin{figure}
\centering
\includegraphics [width=0.5\textwidth]{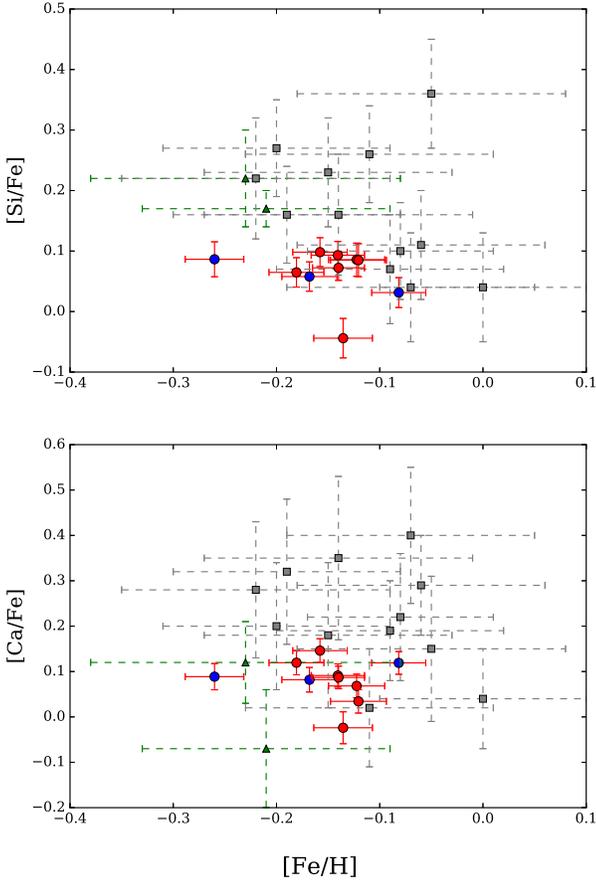} 
\caption{[Si/Fe] and [Ca/Fe] as a function of [Fe/H]. Symbols are explained in Figures \ref{fig:CNONa} and \ref{fig:AlMg}. }\label{fig:alphas}
\end{figure}

We have shown that O and Mg abundances may be modified in the SG stars through the CNO and MgAl cycles. In Figure \ref{fig:alphas}, the two generations of stars have indistinguishable Si and Ca abundances. We note that the warmest star (Star 2) has the lowest [Si/Fe] and [Ca/Fe], probably due to the weaker transition lines in hotter stars. As we show above, the Si precision in APOGEE is generally good, but there seems to be a zeropoint offset between the APOGEE data and the literature. In general, we find no obvious difference in Ca abundances between the two generations, which agrees with the similar statement of \citet{Carretta2010} and \cite{Meszaros2015}. The APOGEE Ca abundances show smaller scatter and smaller errors than the optical Ca abundances.
  
To summarize, we see no clear abundance differences in Si and Ca. Using them as indicators, we infer no significant different contribution of Type II SNe in the FG and SG stars.

\subsection{Iron-peak elements}
\label{sect:iron}
\begin{figure}
\centering
\includegraphics [width=0.5\textwidth]{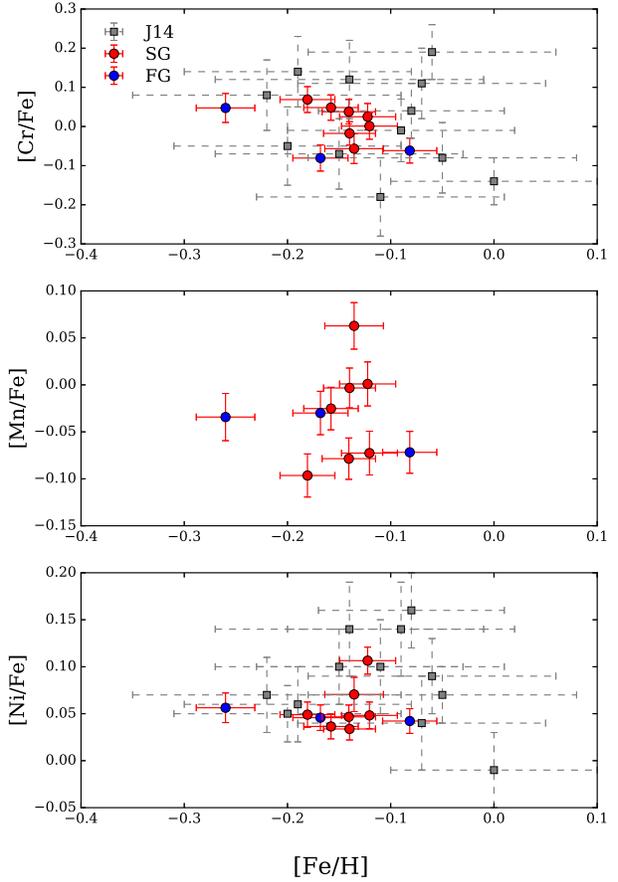} 
\caption{Iron-peak elements as a function of [Fe/H]. Symbols are explained in Figures \ref{fig:CNONa} and \ref{fig:AlMg}. }\label{fig:Fepeak}
\end{figure}

Though Type Ia SNe, runaway deflagration obliterations of white dwarfs, have a signature more tilted towards the iron-peak group \citep{Nomoto1997}, the solar composition of the iron-peak elements are in fact a heterogeneous combination of both Type Ia SNe and core collapse Type II SNe \citep{Woosley1995}. As one of the most metal-rich GCs, our sample stars of NGC 6553 show a mean iron abundance of $-0.15\pm 0.05$. [Cr/Fe], [Mn/Fe], and [Ni/Fe] are mostly within $\pm0.1$ dex of the solar abundances (Figure \ref{fig:Fepeak}).  [Cr/Fe] and [Mn/Fe] from APOGEE also have smaller scatter than their optical counterparts. No chemical difference is found in [Fe/H], [Cr/Fe], [Mn/Fe], and [Ni/Fe] for FG and SG stars. Apart from a few notable iron-complex GCs in the literature, the levels of [Fe/H]  and iron-peak elements are very constant in GCs \citep{Carretta2009iron}. Therefore, it seems that Type Ia SNe do not significantly pollute SG stars.

\section{Discussion}
\label{sect:dis}

\subsection{Elements with larger uncertainties}
\label{sect:other}

In this section, we discuss the elements with larger uncertainties presented in DR13.  
We first notice that FG and SG stars seem to show different abundances in Ti, \tiii, and V.
To test this notion, we run the two sample Kolmogorov$-$Smirnov (KS) test on our sample stars. For [C/Fe],  [\ci/Fe], [N/Fe] (raw), [O/Fe], and [Na/Fe], the KS test shows P$_{\rm r}\leqslant$0.05, which means that there is a $\leqslant5$\% chance that these two samples were drawn from the same parent distribution. Surprisingly, the two generations also show differences in Ti, \tiii, and V with P$_{\rm r}\leqslant$0.05. The reader is reminded that our sample size is rather limited, the KS test should be treated as illustrative, but provocative.
Ti is generally considered as an $\alpha$ element, while V as an iron-peak element.
The different Ti abundances in two generations indicate at face value that a substantial amount of Ti is generated for SG stars, the opposite of O.  However, the Ti and \tiii~ abundances from APOGEE may be subject to large uncertainties. The Ti and \tiii~ abundances may show substantial differences for the same star, e.g., the largest difference is found in the most Na-poor stars of our APOGEE sample (Star 1). Recent study of \citet{Hawkins2016} suggests that some of the \tii~lines may be affected by NLTE effects. Moreover, the two generations of stars also show a separation in V abundances. [V/Fe] is currently not recommended because it displays a large scatter. \citet{Hawkins2016} suggested that V may be similarly affected by NLTE as Ti, since the [V/Fe]$-$[Fe/H] pattern becomes consistent with the literature after their line selection. In addition, the possible temperature systematics from ASPCAP ($\S$\ref{sect:tg}) may drive the low excitation potential features, such as Ti and V, to change oppositely with respect to O.


K, one of the Alkali metals, shows similar abundances for FG and SG stars, which may indicate different nucleosynthesis than the other Alkali metal, Na. This is quite possible since Na has been modified by the NeNa cycle. 
S abundances show a scatter of 1 dex in our sample stars, because its lines are too weak to be properly identified in the APOGEE spectra of our sample stars. For other elements newly identified in DR13 (P, Co, Cu, Ge, and Rb), there are a few things to note: the detection of the Rb line is extremely challenging; Cu and P both present two promising and strong lines, but one of the P lines is in a region heavily affected by telluric features \citep{Hawkins2016}. These elements tend to show large scatter and possible temperature trend (e.g., P in this work) in DR13. So we defer the studies of these elements to a future data release.

\subsection{Comparison with literature works}
\label{sect:liter}

On the basis of the Na-O anti-correlation in 15 GCs, \citet{Carretta2009a} concluded that the FG stars in GCs amounts to about 30\% of the total population, what they call intermediate population amounts to almost 60\% and in some clusters there is an extra about 10\% in an extreme population (a similar conclusion has been obtained by \citealt{Bastian2015}).
Using the APOGEE data, we clearly separate two generations of stars, and the FG stars account for 30\% of the sample, which is similar to the Carretta et al. value. However, we do not claim this to be statistically significant, due to the limited sample size of this work and unknown selection effects. A larger sample is required.

\citet{Carretta2009b} found that the Al$-$Mg anti-correlation and Al$-$Si correlation tend to be more significant in metal-poor or massive GCs, while Ca does not show discernible difference between two generations in any GC \citep{Carretta2010}. \cite{Meszaros2015} also drew similar conclusions on the aforementioned elements with a self-consistent study of 10 GCs observed by APOGEE.
NGC 6553 stars do not show a clear Al$-$Mg anti-correlation, and no distinguishable Si (and Ca) difference can be found for the two generations. As a metal-rich and intermediate-mass GC, NGC 6553 stars qualitatively agree with the literature results.

\subsection{AGB polluting models}

Historically, AGB stars \citep{Ventura2011, Ventura2013}, fast rotating massive stars \citep{Decressin2007},  super-massive stars \citep{Denissenkov2014}, and massive interactive binaries \citep{deMink2009} have been proposed to predict the chemical behaviors of observed GCs. In this work, we select the AGB polluting models from \citet[][V13]{Ventura2013} to compare with our observations, because  \citet{Renzini2015} argued that: (1) scenarios appealing to super-massive stars, fast rotating massive stars and massive interactive binaries violate in an irreparable fashion two or more constraints among their seven observational constraints; (2) the AGB models are not totally consistent with observational constraints (e.g., mass budget problem), but there seem to be ways to save it; (3) the AGB models explicitly illustrate the metallicity and mass dependence of the observed correlations. 
In the AGB polluting scheme, the dependence on metallicity for various correlations is a reflection of the fact that the Hot-Bottom Burning (HBB) is expected to occur at a higher temperature in more metal-poor stars. Meanwhile, the high-mass AGB stars reach higher temperature at the bottom of the convective envelope, i.e., stronger HBB. Therefore, the advanced nucleosynthetic processes (e.g., MgAl cycle and Si leakage) tend to occur in GCs with low metallicity or in GCs where pollution from massive AGB stars occurred.

The improved models of \citet{Ventura2011, Ventura2013} presented yields from stars of mass in the range 1 M$_{\odot}\leq$ M $\leq 8$ M$_{\odot}$ of metallicities Z$=3\times10^{-4}$, $10^{-3}$ and $8\times10^{-3}$. The Al$-$Mg anti-correlation can be qualitatively reproduced in the metal-poor models, but an extra dilution mechanism between gas ejection and primordial material may be required to reproduce the observed Na$-$O anti-correlation.
Here we confront the metal-rich AGB models with our observations. In the V13 metal-rich models, stars with initial stellar mass lower than 3 M$_{\odot}$ are dominated by third dredge-up (TDU), while HBB starts to take control of the evolution at the threshold mass of 3.5 M$_{\odot}$. At the latter phase, the maximum temperature reached by the bottom of the convective envelope jumps to more than 80 MK. This temperature is very important, since various proton capture channels require $\sim100$ MK to be activated in the V13 models. 

Pure AGB yields from the most metal-rich V13 models are plotted as cyan stars in Figures \ref{fig:CNONa}, \ref{fig:AlMg}, \ref{fig:SiAl}, \ref{fig:AlNa}, and \ref{fig:AlO}. The primordial abundances (P) are set to solar for C, N, Na, and Al, but $+0.2$ for O, Mg, and Si to account for the $\alpha$ enhancement. Initial stellar masses of the polluting AGB stars are 3.5, 4, 4,5, 5.0, 5.5, 6.0, 6.5, 7.0, 7.5, and 8.0 M$_{\odot}$. We indicate 3.5, 8.0 M$_{\odot}$, and sometimes 5.5 M$_{\odot}$ in the figures, since the last model shows the highest [Al/Fe]. The iron abundance of these models is in fact $-0.5$, somewhat more metal-poor than that of NGC 6553. 
However, the recent extension of the massive AGB V13 models to solar metallicity
\citep{DiCriscienzo2016} suggests that the AGB yields of the key elements
(e.g., N, Na) do not vary significantly between [Fe/H]$\sim-0.5$ (V13)
and solar metallicity (see e.g., their Figures 9 and 10).

In Figure \ref{fig:CNONa}, the FG stars show similar [C/Fe], [N/Fe], [O/Fe], and [Na/Fe] as the primordial abundances, suggesting that the initial abundances determined by the chemical evolution of the Galaxy are also valid in NGC 6553. The SG stars show consistent [N/Fe] and [O/Fe] with the models, though the readers are reminded that the observational N abundances are estimated near the synthetic spectrum grid edge. 
In the V13 AGB models, the C depletion and Na enhancement are clearly stronger
than those measured in the most contaminated stars in the cluster. 
The data of NGC 6553 can only be matched under the AGB scenario if
some dilution of the AGB ejecta with pristine material is assumed to happen.  Such 
dilution is required to explain both the Na-O anti-correlation and to reduce the
discrepancy between model predictions and the data for [C/Fe] and [Na/Fe]. Figure
8 shows that one FG star has large [Al/Fe], which is close to the SG stars, but
the other two FG stars have [Al/Fe] abundances similar to the primordial values.
The [Mg/Fe] values also match reasonably well for FG stars. The smaller scatter
observed in [Mg/Fe] is well reproduced by the almost constant [Mg/Fe] in the
pure AGB models, but the dynamic range of [Al/Fe] in the models is slightly
smaller (although only of the order of the abundance errors of about 0.1 dex)
than the observed one\footnote{\citet{Ventura2011} attempted to fix the small
[Al/Fe] dynamic range problem in metal-poor models by increasing the
cross-section of the MgAl chain by a factor of 2 with respect to the highest
value allowed by the European compilation of reaction rates for astrophysics compilation. This revision reproduces the dynamic
ranges of Mg depletion and Al enhancement observed in metal-poor GCs. Therefore,
it may be worth investigating the similar subject in the more metal-rich
models.}. In this case, the dilution correction has a much smaller impact on
the Al abundances. For example, the same degree of pollution required to
decrease the Na of the ejecta (by $\sim0.7$ dex) to the highest value observed
($\sim0.4-0.5$ dex) would hardly change the situation for Al, as the expected
abundances of the latter element would decrease from $\sim0.3$ dex (in the pure AGB
ejecta) to $\sim0.2$ dex (after dilution correction) (P. Ventura, private
comm.).

The observed [Si/Fe] in Figure \ref{fig:SiAl} obviously show an offset with respect to the models, which may be again related with the zeropoint issue found in DR13 calibrated Si abundances. However, the observed constant [Si/Fe] behavior is well reproduced by the models. This implies that the Si leakage from the MgAl chain is not present or very weak in NGC 6553. In Figures \ref{fig:AlNa} and \ref{fig:AlO}, we again see that the theoretical primordial abundances are representative of the FG stars, but the slopes of the observed correlations are different than the models, mainly due to the lower [Al/Fe] and higher [Na/Fe] produced in the pure AGB models. 


Before ending this section, there are some caveats in modeling that we should note. We compare the pure model yield predictions directly with the stellar abundance data for GC members following \citet{Ventura2011}. These comparisons are only meaningful under the notion that SG stars with a given enhanced abundance pattern are only enriched (or depleted) by those FG stars in the mass range required to generate the needed yields.  An exceedingly fine tuning of the star formation and chemical evolution time scales is needed in order to accomplish even rough agreement with the observations.  Such somewhat contrived requirements are posed by all self-enrichment models, so they are not unique to the particular models adopted for the comparisons in this work. Under those assumptions, our comparison with the pure AGB metal-rich V13 models shows that data can only be matched by assuming some dilution of the AGB ejecta with pristine material in order to reproduce the observed Na-O anti-correlation and the [C/Fe], [Na/Fe], and [Al/Fe] abundances. We defer a detailed quantitative confrontation between model and data to a future publication.

\section{Summary}
\label{sect:con}

We present our study of the stellar chemical abundances of the bulge GC NGC 6553 using calibrated APOGEE values from SDSS DR13. Ten red giants are identified as cluster members using their positions, radial velocities, iron abundances, and NIR photometry. Our sample stars have a mean RV of $-0.14\pm5.47$ km s$^{-1}$, and a mean [Fe/H] of $-0.15\pm 0.05$, which are consistent with the more recent literature results. We clearly separate two populations of stars in C and N in this GC for the first time. Three stars are grouped as first generation (FG), while seven stars are grouped as second generation (SG). Strong N$-$C, Na$-$O anti-correlations  are found with the abundances derived by APOGEE. The Al$-$Na correlation and Al$-$O anti-correlation are also confirmed. However, the Mg$-$Al anti-correlation cannot be confirmed since [Mg/Fe] show a scatter comparable to that from ASPCAP. We see no obvious difference in Si and Ca abundances for the two generations. Therefore, our results suggest that the CNO, NeNa, MgAl cycles have been activated, but the MgAl cycle is too weak to show its effect in Mg. The Si leakage from the MgAl cycle is also weak. Two generations show similar  iron-peak element abundances (Fe, Cr, Mn, and Ni), which suggests that Type Ia SNe do not significantly affect the pollution of SG stars. 

As a metal-rich and intermediate-mass GC, our results in NGC 6553 generally agree with the current knowledge about GC stellar abundance correlations, i.e., weaker Al$-$Mg correlation and no Si (or Ca) variation in metal-rich GCs.
We also compare our results with the AGB polluting models of \citet{Ventura2013}. Our comparison with the pure AGB metal-rich V13 models shows that the observed data can only be matched by assuming some dilution of the AGB ejecta with pristine material. In addition, self-enrichment models with star formation and chemical evolution is necessary for further GC chemical studies.


From a sample of ten cluster members, we unequivocally identified two groups of stars with distinct chemical abundances in NGC 6553. Meanwhile, our discussions about various correlations (e.g., the Al$-$Na correlation) should be confirmed by a larger, self-consistent sample. Detailed studies of chemical abundances in other bulge GCs (Schiavon et al. 2016, submitted) will be also very helpful to understand the astrophysical processes in these stellar systems. 

\section{acknowledgments}

We thank F. Mauro, P. Ventura and J. Holtzman for helpful discussions, and
the anonymous referee for insightful comments. D.G., B.T. and S.V. gratefully acknowledges support from the Chilean BASAL   Centro de Excelencia en Astrof\'{i}sica y Tecnolog\'{i}as Afines (CATA) grant PFB-06/2007.
R.E.C. acknowledges funding through Gemini-CONICYT Project 32140007.
D.A.G-H was funded by the Ramon y Cajal fellowship number RYC-2013-14182. D.A.G-H and
O.Z. acknowledge support provided by the Spanish Ministry of Economy and
Competitiveness (MINECO) under grant AYA-2014-58082-P.
A.M. acknowledges support from Proyecto Interno UNAB DI-677-15/N.
J.G.F-T is currently supported by Centre National d'Etudes Spatiales (CNES) through PhD grant 0101973 and the R\'egion de Franche-Comt\'e and by the French Programme National de Cosmologie et Galaxies (PNCG).

Funding for the Sloan Digital Sky Survey IV has been provided by the
Alfred P. Sloan Foundation, the U.S. Department of Energy Office of
Science, and the Participating Institutions. SDSS- IV acknowledges
support and resources from the Center for High-Performance Computing at
the University of Utah. The SDSS web site is www.sdss.org.

SDSS-IV is managed by the Astrophysical Research Consortium for the Participating Institutions of the SDSS Collaboration including the Brazilian Participation Group, the Carnegie Institution for Science, Carnegie Mellon University, the Chilean Participation Group, the French Participation Group, Harvard-Smithsonian Center for Astrophysics, Instituto de Astrof\`{i}sica de Canarias, The Johns Hopkins University, Kavli Institute for the Physics and Mathematics of the Universe (IPMU) / University of Tokyo, Lawrence Berkeley National Laboratory, Leibniz Institut f\"{u}r Astrophysik Potsdam (AIP), Max-Planck-Institut f\"{u}r Astronomie (MPIA Heidelberg), Max-Planck-Institut f\"{u}r Astrophysik (MPA Garching), Max-Planck-Institut f\"{u}r Extraterrestrische Physik (MPE), National Astronomical Observatory of China, New Mexico State University, New York University, University of Notre Dame, Observat\'{o}rio Nacional / MCTI, The Ohio State University, Pennsylvania State University, Shanghai Astronomical Observatory, United Kingdom Participation Group, Universidad Nacional Aut\'{o}noma de M\'{e}xico, University of Arizona, University of Colorado Boulder, University of Oxford, University of Portsmouth, University of Utah, University of Virginia, University of Washington, University of Wisconsin, Vanderbilt University, and Yale University.

\begin{table*}
\caption{DR13 calibrated abundances and errors for NGC 6553 cluster members.}              
\label{tab2}      
\setlength{\tabcolsep}{2pt} 
\centering                                      
\begin{tabular}{c c c c c c c c c c c c c c c c c}          
\hline\hline                        
  \#      &     [C/Fe] & $\delta_{\rm [C/Fe]}$& [\ci/Fe] & $\delta_{[\rm C\,I/Fe]}$ &[N/Fe]  &$\delta_{\rm [N/Fe]}$ &  [O/Fe]  &$\delta_{\rm [O/Fe]}$&  [Na/Fe] &$\delta_{\rm [Na/Fe]}$ &  [Mg/Fe] &$\delta_{\rm [Mg/Fe]}$  &  [Al/Fe]&$\delta_{\rm [Al/Fe]}$ &   [Si/Fe]  &  $\delta_{\rm [Si/Fe]}$    \\

\hline
1 &  0.13 &  0.03 &  0.12 &  0.05 &  0.24 &  0.06 &  0.14 &  0.02 & -0.02 &  0.08 &  0.18 &  0.02 & 0.00 &  0.07 &  0.03 &  0.02 \\
2 & -0.26 &  0.04 & -0.35 &  0.06 & -- & -- &  0.03 &  0.05 &  0.34 &  0.09 &  0.13 &  0.02 &  0.32 &  0.08 & -0.04 &  0.03 \\
3 & -0.12 &  0.03 & -0.09 &  0.05 & -- & -- &  0.12 &  0.02 &  0.46 &  0.09 &  0.14 &  0.02 &  0.36 &  0.07 &  0.06 &  0.02 \\
4 & -0.22 &  0.03 & -0.22 &  0.05 & -- & -- &  0.12 &  0.02 &  0.41 &  0.08 &  0.15 &  0.02 &  0.37 &  0.07 &  0.09 &  0.02 \\
5 & -0.17 &  0.03 & -0.17 &  0.04 &  0.88 &  0.05 &  0.09 &  0.01 &  0.34 &  0.08 &  0.14 &  0.02 &  0.24 &  0.07 &  0.07 &  0.02 \\
6 & -0.31 &  0.03 & -0.29 &  0.05 & -- & -- &  0.07 &  0.03 &  0.28 &  0.09 &  0.14 &  0.02 &  0.17 &  0.07 &  0.09 &  0.03 \\
7 & -0.26 &  0.03 & -0.24 &  0.05 & -- & -- &  0.17 &  0.02 &  0.44 &  0.09 &  0.17 &  0.02 &  0.26 &  0.07 &  0.10 &  0.02 \\
8 &  0.13 &  0.04 &  0.15 &  0.06 &  0.38 &  0.07 &  0.21 &  0.03 &  0.20 &  0.10 &  0.18 &  0.02 & -0.09 &  0.07 &  0.09 &  0.03 \\
9 &  0.07 &  0.03 &  0.06 &  0.05 &  0.29 &  0.06 &  0.14 &  0.02 &  0.05 &  0.09 &  0.18 &  0.02 &  0.18 &  0.07 &  0.06 &  0.02 \\
10 & -0.31 &  0.03 & -0.30 &  0.05 & -- & -- &  0.12 &  0.03 &  0.29 &  0.08 &  0.12 &  0.02 &  0.19 &  0.07 &  0.08 &  0.03 \\
\hline 
  \#      &     [P/Fe] & $\delta_{\rm [P/Fe]}$& [S/Fe] & $\delta_{\rm [S/Fe]}$ &[K/Fe]  &$\delta_{\rm [K/Fe]}$ &  [Ca/Fe]  &$\delta_{\rm [Ca/Fe]}$&  [Ti/Fe] &$\delta_{\rm [Ti/Fe]}$ &  [\tiii/Fe] &$\delta_{\rm [Ti\,II/Fe]}$  &  [V/Fe]&$\delta_{\rm [V/Fe]}$ &   [Cr/Fe]  &  $\delta_{\rm [Cr/Fe]}$    \\
\hline 
1 &  0.17 &  0.11 &  0.17 &  0.10 &  0.08 &  0.03 &  0.12 &  0.03 &  0.12 &  0.04 & -0.27 &  0.09 &  0.13 &  0.05 & -0.06 &  0.03 \\
2 &  0.47 &  0.13 & -0.24 &  0.12 &  0.28 &  0.05 & -0.02 &  0.04 &  0.26 &  0.05 &  0.46 &  0.11 &  0.07 &  0.09 & -0.06 &  0.04 \\
3 &  0.17 &  0.11 &  0.14 &  0.10 &  0.12 &  0.04 &  0.12 &  0.03 &  0.25 &  0.04 &  0.17 &  0.09 &  0.17 &  0.05 &  0.07 &  0.03 \\
4 &  0.20 &  0.10 & -0.22 &  0.10 &  0.24 &  0.03 &  0.09 &  0.03 &  0.28 &  0.04 &  0.37 &  0.09 &  0.26 &  0.05 &  0.04 &  0.03 \\
5 &  0.12 &  0.09 & -0.13 &  0.10 &  0.17 &  0.03 &  0.09 &  0.02 &  0.34 &  0.04 &  0.27 &  0.08 &  0.26 &  0.04 & -0.02 &  0.03 \\
6 &  0.22 &  0.11 &  0.04 &  0.10 &  0.14 &  0.04 &  0.07 &  0.03 &  0.28 &  0.04 &  0.18 &  0.10 &  0.19 &  0.06 &  0.02 &  0.03 \\
7 &  0.22 &  0.10 & -0.06 &  0.10 &  0.02 &  0.03 &  0.15 &  0.03 &  0.34 &  0.04 &  0.22 &  0.09 &  0.36 &  0.05 &  0.05 &  0.03 \\
8 &  0.40 &  0.12 &  0.50 &  0.10 &  0.00 &  0.04 &  0.09 &  0.03 &  0.01 &  0.05 &  0.07 &  0.10 & -0.04 &  0.07 &  0.05 &  0.04 \\
9 & -0.04 &  0.11 &  0.03 &  0.10 &  0.16 &  0.04 &  0.08 &  0.03 &  0.08 &  0.04 &  0.16 &  0.09 & -0.02 &  0.05 & -0.08 &  0.03 \\
10 &  0.12 &  0.11 &  0.31 &  0.10 &  0.13 &  0.04 &  0.03 &  0.03 &  0.19 &  0.04 &  0.13 &  0.10 &  0.16 &  0.06 &  0.00 &  0.03 \\
\hline 
\end{tabular}
\begin{tabular}{c c c c c c c c c c c c c c c }          
  \#      &     [Mn/Fe] & $\delta_{\rm [Mn/Fe]}$& [Co/Fe] & $\delta_{\rm [Co/Fe]}$ &[Ni/Fe]  &$\delta_{\rm [Ni/Fe]}$ &  [Cu/Fe]  &$\delta_{\rm [Cu/Fe]}$&  [Ge/Fe] &$\delta_{\rm [Ge/Fe]}$ &  [Rb/Fe] &$\delta_{\rm [Rb/Fe]}$   &  [N/Fe](raw) &$\delta_{\rm [N/Fe](raw)}$   \\
\hline 
1 & -0.07 &  0.02 &  0.20 &  0.07 &  0.04 &  0.01 & -0.04 &  0.10 & -0.19 &  0.15 &  0.04 &  0.08 &  0.24 &  0.01 \\
2 &  0.06 &  0.02 &  0.43 &  0.13 &  0.07 &  0.02 &  0.20 &  0.09 & -0.05 &  0.17 & -0.45 &  0.18 &  0.94 &  0.03 \\
3 & -0.10 &  0.02 &  0.04 &  0.07 &  0.05 &  0.01 & -- & -- &  0.04 &  0.12 &  0.08 &  0.08 &  1.00 &  0.00 \\
4 & -0.08 &  0.02 &  0.17 &  0.06 &  0.05 &  0.01 &  0.02 &  0.11 &  0.02 &  0.12 & -0.11 &  0.08 &  1.00 &  0.00 \\
5 & -0.00 &  0.02 &  0.22 &  0.05 &  0.03 &  0.01 &  0.16 &  0.11 &  0.03 &  0.11 & -0.00 &  0.07 &  0.90 &  0.00 \\
6 &  0.00 &  0.02 &  0.45 &  0.09 &  0.11 &  0.01 &  0.21 &  0.10 & -0.18 &  0.17 & -0.05 &  0.09 &  0.98 &  0.01 \\
7 & -0.03 &  0.02 & -0.07 &  0.07 &  0.04 &  0.01 &  0.07 &  0.11 &  0.07 &  0.12 & -0.00 &  0.08 &  1.00 &  0.00 \\
8 & -0.03 &  0.03 &  0.04 &  0.10 &  0.06 &  0.02 &  0.04 &  0.11 &  0.11 &  0.16 & -0.36 &  0.14 &  0.39 &  0.01 \\
9 & -0.03 &  0.02 & -0.03 &  0.07 &  0.05 &  0.01 &  0.06 &  0.11 &  0.32 &  0.15 &  0.06 &  0.08 &  0.29 &  0.01 \\
10 & -0.07 &  0.02 &  0.00 &  0.10 &  0.05 &  0.01 &  0.11 &  0.10 & -0.04 &  0.16 & -0.08 &  0.09 &  0.96 &  0.01 \\
\hline 
\end{tabular}
\raggedright{{\bf Note:} All values are given in the unit of dex. [N/Fe] (raw) are not calibrated. See text for information about [N/Fe] (raw).}
\end{table*}

\bibliographystyle{mn}
\bibliography{gc,apo}

\label{lastpage}

\end{document}